**Analytical modeling of laminated composite rings on nonreciprocal elastic foundations under non-axisymmetric loading**

Zhipeng Liu, Jaehyung Ju*

UM-SJTU Joint Institute, Shanghai Jiao Tong University

800 Dongchuan Road, Shanghai 200240, China

**Abstract**

A mechanical model of a laminated composite ring on a nonreciprocal elastic foundation is a valuable engineering tool during the early design stages of various applications, such as non-pneumatic wheels, flexible bearings/bushings, expandable tubulars in oil wells, and vascular stents interacting with blood vessel linings, especially under non-axisymmetric loadings. Despite its importance, limited research has focused on the interaction between laminated composite rings and nonreciprocal elastic foundations. Moreover, no quantitative studies have yet explored the influence of foundation stiffness on the ring's deformation. This work aims to develop an analytical framework for a laminated composite ring supported by a nonreciprocal elastic foundation under non-axisymmetric loading conditions. The model generates a design map that correlates the foundation's stiffness with the ring's deformation, accounting for ring dimensions, laminate lay-up architecture, and lamina anisotropy. The closed-form solution provides an efficient design tool for analyzing non-axisymmetric and nonuniform loadings at a low computational cost. The resulting design map provides a valuable resource for exploring the interaction between the nonreciprocal foundation and the laminated ring. The proposed analytical framework and design map hold broad potential applications in automotive, mechanical, civil, and biomedical engineering fields.

## 1. Introduction

The flexible ring on an elastic foundation (REF) model is widely used in engineering to analyze the mechanical behavior of rings supported by elastic media. Applications include non-pneumatic wheels with shear bands supported by elastic spokes [1], pipelines resting on soil foundations [2], gears [3], pneumatic tires [4], and railway wheels [5]. This model provides valuable insights into the interaction of the ring with external forces and moments [6], [7], [8].

In many engineering applications, rings are designed using advanced composite materials and layered architectures to optimize performance in terms of weight and structural efficiency. For example, tires often consist of multiple layers, including tread, belts, and carcasses, each engineered for specific properties such as traction, strength, and flexibility [9]. Similarly, multi-layered pipelines enhance load-bearing capacity and resistance to environmental factors [10]. However, traditional models that treat rings as homogenized entities [11], [12], [13], [14], [15] may fail to capture the complex interactions and mechanical behaviors of individual layers, potentially leading to less accurate predictions of stress and deformation. Therefore, it is essential to develop laminated composite ring models that provide more precise stress distribution and deformation predictions under various loading conditions than homogenized models [16], [17].

* Corresponding author: jaehyung.ju@sjtu.edu.cn

Many existing models, based on classical laminated plate theory [18], simplify the analysis by assuming plane stress conditions and neglecting interlayer shear effects [19], [20], [21]. However, recent advancements in higher-order shear deformation theories [21], [22], [23], [24] and the use of shear correction factors [25] have significantly enhanced the understanding of multi-layered ring mechanics, leading to more accurate predictions of stress distributions and deformations.

Despite these advancements, the study of laminated rings on elastic foundations remains relatively underexplored. Most research has either focused on the behavior of laminated rings without considering interaction with elastic foundations [26], [27], [28], [29], [30], or on homogenized rings interacting with elastic foundations [31], [32], [33], [34]. Additionally, a ring on a nonreciprocal elastic foundation can undergo non-axisymmetric deformation, which is relevant to numerous engineering applications, such as tires, pipelines, non-pneumatic wheels [35], [36], flexible bearings and bushings [37], expandable tubulars in oil wells [38], and vascular stents interacting with blood vessel linings [39].

Therefore, this work aims to develop an analytical framework for a laminated composite ring on a nonreciprocal elastic foundation under non-axisymmetric loading conditions. We present a closed-form solution for the laminated ring on an elastic foundation model subjected to radial loads in one direction, enabling accurate predictions of the overall deformation patterns in the ring-foundation system. The model also generates a design map that correlates the foundation's stiffness with the ring's deformation, accounting for ring dimensions, laminate lay-up architecture, and lamina anisotropy.

## 2. Laminated curved beam theory

We consider an elastic laminated curved beam where adjoining layers are perfectly bonded with no relative sliding or debonding. Figure 1a illustrates the geometry of a laminated curved beam characterized by a midplane radius $R$ and a rectangular cross-section with a thickness $h$ and consistent width $b$. Based on the first-order laminated plate theory [25], the radial and circumferential displacements in polar coordinates, $s_0$, for the laminated curved beam can be determined by

$$\begin{aligned} u_r(r,\theta) &= u_r(\theta) \\ u_\theta(r,\theta) &= u_{\theta 0}(\theta) + z\phi(\theta) \end{aligned} \tag{1}$$

where $u_r(r,\theta)$ and $u_\theta(r,\theta)$ denote the radial and circumferential displacements at a point of the curved beam. Notably, the radial displacement is the only function of the circumferential position $\theta$. $u_{\theta 0}(\theta)$ and $\phi(\theta)$ are the circumferential displacement and the cross-sectional rotation along the midplane of the beam, respectively. $z$ represents a position along the thickness measured positively outward from the curved beam's midplane, as shown in Figure 1a. Given the small deformation of $z/R$ terms [21], the circumferential strain and shear strain ($\varepsilon_{\theta\theta}$ and $\gamma_{r\theta}$) at any position $z$ can be expressed as

$$\varepsilon_{\theta\theta} = \frac{1}{1+z/R}\left(\varepsilon_{\theta\theta}^{0} + z\kappa\right)$$
$$\gamma_{r\theta} = \frac{1}{1+z/R}\gamma_{r\theta}^{0} \tag{2}$$

where $\varepsilon_{\theta\theta}^{0}$, $\gamma_{r\theta}^{0}$ and $\kappa$ represent the circumferential and shear strains and curvature change in the middle surface, respectively, which are given by

$$\varepsilon_{\theta\theta}^{0} = \frac{1}{R}\left(u_r + \frac{du_{\theta 0}}{d\theta}\right)$$
$$\kappa = \frac{1}{R}\frac{d\phi}{d\theta} \tag{3}$$
$$\gamma_{r\theta}^{0} = \frac{1}{R}\left(\frac{du_r}{d\theta} - u_{\theta 0}\right) + \phi$$

The stress-strain relation for a material element of the lamina in the $k$th layer is expressed as [40],

$$[\boldsymbol{\sigma}]_k = [\bar{\boldsymbol{Q}}]_k[\boldsymbol{\varepsilon}]_k \tag{4}$$

Here $\boldsymbol{\sigma}$ is the stress tensor, $\boldsymbol{\varepsilon}$ is the strain tensor, and $\bar{\boldsymbol{Q}}$ is the transformed reduced stiffness tensor of the $k$th layer. For a laminated curved beam with $n$ layers, the internal axial force ($N$), moment ($M$), and shear force ($V$) can be obtained by the sum of integrals of stresses over the thickness of each layer, which are

$$[N, M, V] = b\sum_{k=1}^{N}\int_{z_{k-1}}^{z_k}[\sigma_{\theta\theta}, z_n\sigma_{\theta\theta}, \tau]\,dz \tag{5}$$

Inserting Eq. (4) into Eq. (5) yields

$$\begin{bmatrix} N \\ M \\ V \end{bmatrix} = \begin{bmatrix} A_{11} & B_{11} & 0 \\ B_{11} & D_{11} & 0 \\ 0 & 0 & A_{55} \end{bmatrix}\begin{bmatrix} \varepsilon_{\theta\theta}^{0} \\ \kappa \\ \gamma_{r\theta}^{0} \end{bmatrix} \tag{6}$$

where the coefficients $A_{11}$, $B_{11}$, and $D_{11}$ denote the extension stiffness, extension and bending coupling stiffness, and bending stiffness arising from the integration over the layer thickness, calculated as

$$A_{11} = Rb\sum_{k=1}^{N}\bar{Q}_{11}^{(k)}\ln\left(\frac{R+z_k}{R+z_{k-1}}\right)$$
$$B_{11} = Rb\sum_{k=1}^{N}\bar{Q}_{11}^{(k)}\left[(z_k - z_{k-1}) - R\ln\left(\frac{R+z_k}{R+z_{k-1}}\right)\right] \tag{7}$$
$$D_{11} = Rb\sum_{k=1}^{N}\bar{Q}_{11}^{(k)}\left[\frac{1}{2}\left((R+z_k)^2 - (R+z_{k-1})^2\right) - 2R(z_k - z_{k-1}) + R^2\ln\left(\frac{R+z_k}{R+z_{k-1}}\right)\right]$$

where $z_k$ is the distance from the midplane to the surface of the $k$th layer. And the shearing stiffness $A_{55}$ is calculated as

$$A_{55} = b\sum_{k=1}^{N}\int_{z_{k-1}}^{z_k}\frac{K_S Q_{55}^{(k)}}{1+z/R}dz \tag{8}$$

$K_S$ represents the shear correction coefficient [25], which corrects the discrepancy between the constant stress state predicted by the first-order laminated plate theory and the actual stress state, where the transverse shear stresses are distributed at least quadratically across the laminate thickness.

Considering a parabolic distribution of the shear stress, which is defined as [40]

$$K_S(z) = \frac{5}{4}\left[1 - \left(\frac{z}{h/2}\right)^2\right] \tag{9}$$

Inserting Eq. (9) into Eq. (8), we can rewrite the shearing stiffness $A_{55}$ in Eq. (8) as

$$A_{55} = \frac{5}{4} bR \sum_{k=1}^{N} \bar{Q}_{55}^{(k)} \left[ -\frac{2}{h^2}\left(z_k^2 - z_{k-1}^2\right) + \frac{4R}{h^2}(z_k - z_{k-1}) + \left(1 - \frac{4R^2}{h^2}\right) \ln\left(\frac{R + z_k}{R + z_{k-1}}\right) \right] \tag{10}$$

Figure 1b illustrates a free-body diagram of axial forces ($N$), shear forces ($V$), moments ($M$), and the radial and circumferential distributed loads, $q_r$ and $q_\theta$, acting on an infinitesimal element of a laminated curved beam with the neutral surface length $ds_0$ ($= Rd\theta$). $q_r$ and $q_\theta$ are applied on the middle surface of the beam, and their units are in force per area. The force and moment equilibrium in the infinitesimal element is given by

$$\begin{aligned} &\frac{dN}{Rd\theta} + \frac{V}{R} + bq_\theta = 0 \\ &\frac{N}{R} - \frac{dV}{Rd\theta} - bq_r = 0 \\ &\frac{dM}{Rd\theta} - V = 0 \end{aligned} \tag{11}$$

Substituting Eqs. (3) and (6) into Eq. (11) gives the equations of motion in terms of the radial displacement, circumferential displacement, and the cross-sectional rotation along the middle surface of the laminated curved beam:

$$\begin{aligned} &\frac{A_{11}}{R^2}\frac{d^2u_{\theta 0}}{d\theta^2} - \frac{A_{55}}{R^2}u_{\theta 0} + \frac{A_{11} + A_{55}}{R^2}\frac{du_r}{d\theta} + \frac{B_{11}}{R^2}\frac{d^2\phi}{d\theta^2} + \frac{A_{55}}{R}\phi = -bq_\theta \\ &\frac{A_{11} + A_{55}}{R^2}\frac{du_{\theta 0}}{d\theta} - \frac{A_{55}}{R^2}\frac{d^2u_r}{d\theta^2} + \frac{A_{11}}{R^2}u_r + \frac{B_{11} - RA_{55}}{R^2}\frac{d\phi}{d\theta} = bq_r \\ &\frac{B_{11}}{R^2}\frac{d^2u_{\theta 0}}{d\theta^2} + \frac{A_{55}}{R}u_{\theta 0} + \frac{B_{11} - RA_{55}}{R^2}\frac{du_r}{d\theta} + \frac{D_{11}}{R^2}\frac{d^2\phi}{d\theta^2} - A_{55}\phi = 0 \end{aligned} \tag{12}$$

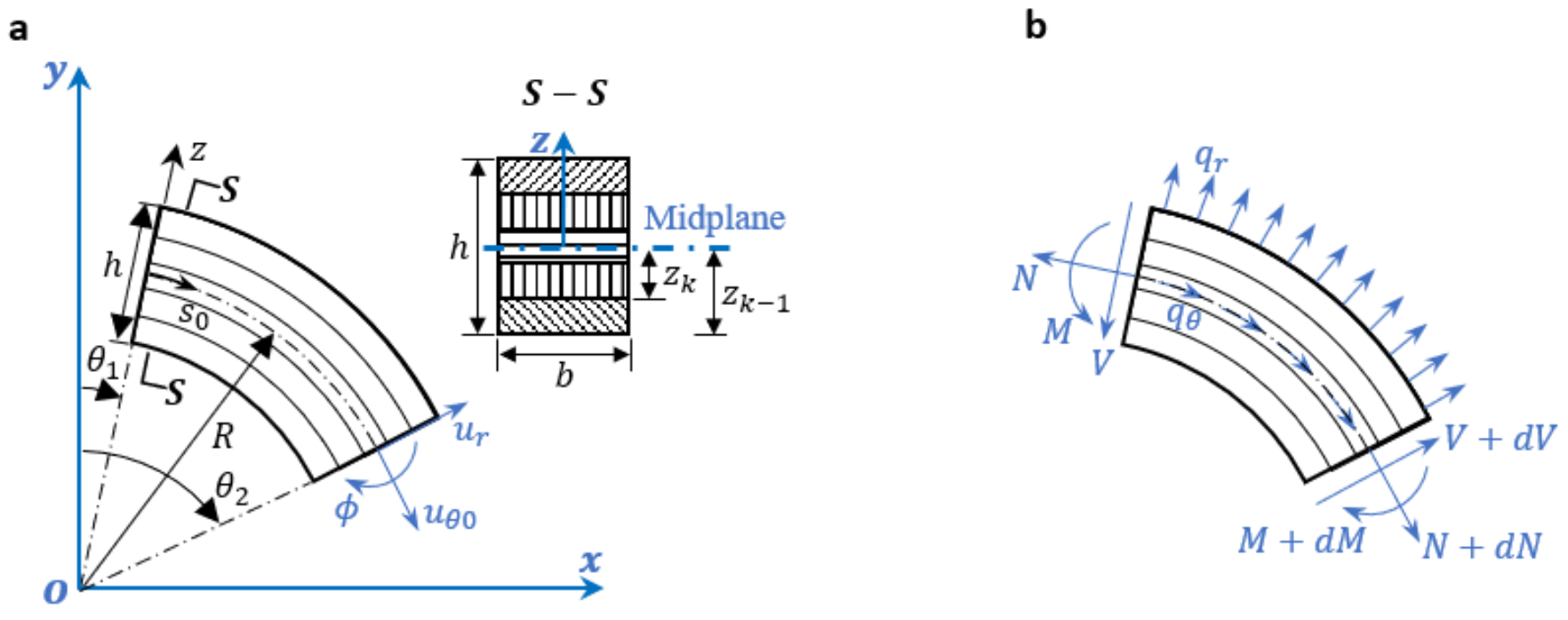

*Figure 1. A laminated curved beam: (a) geometric parameters and displacement fields in the curved beam; (b) force and moment resultants in the curved beam.*

## 3. Laminated ring on an elastic foundation model

Recently, ring models subjected to non-axisymmetric loadings have received significant attention,

particularly in the design of non-pneumatic wheels [12], [13], [14]. Additionally, the nonreciprocal stiffness design of spokes, which improves the load-bearing capacity of wheels, has gained importance in developing modern non-pneumatic wheels, as it ensures better load distribution and enhances ride comfort [1], [36]. In this section, we model a laminated composite ring on an elastic foundation for both point and distributed non-axisymmetric loadings. The linear elastic foundation will be extended to a nonreciprocal form in Section 3.3.

3.1. A laminated ring on a linear elastic foundation (LRLF) subjected to a radial point load

Figure 2 shows a laminated ring where the ring interacts with a linear elastic foundation via an inner radial region. The elastic foundation produces a distributed load per width as $q_r = -k_r u_r(\theta)/b$, where $k_r$ is the radial stiffness of the foundation [41], developing a continuous reaction to the laminated ring. Assuming negligible tangential stiffness of the elastic foundation ($q_\theta = 0$), we insert $q_r = -k_r u_r(\theta)/b$ into Eq. (12) and decouple the equation in terms of $u_r$, having the following expression:

$$\begin{aligned}
&\frac{d^5 u_r}{d\theta^5} + \lambda_1 \frac{d^3 u_r}{d\theta^3} + \lambda_2 \frac{du_r}{d\theta} = 0 \\
&\frac{du_{\theta 0}}{d\theta} = \frac{1}{P(RA_{11} + B_{11})} \frac{d^4 u_r}{d\theta^4} + \frac{(B_{11}P + 1)A_{55} - k_r R^2}{PA_{55}(RA_{11} + B_{11})} \frac{d^2 u_r}{d\theta^2} - \frac{k_r R^3 + RA_{11}}{RA_{11} + B_{11}} u_r \\
&\phi = \frac{1}{R} u_{\theta 0} - \frac{1}{R}\left[1 + \frac{A_{11}D_{11} - B_{11}^2}{A_{55}(RB_{11} + D_{11})}\right] \frac{du_r}{d\theta} - \frac{A_{11}D_{11} - B_{11}^2}{RA_{55}(RB_{11} + D_{11})} \frac{d^2 u_{\theta 0}}{d\theta^2}
\end{aligned} \tag{13}$$

where the constant $P$ is defined as

$$P = \frac{A_{55}}{B_{11} - RA_{55}} \left( \frac{1}{A_{55}} + \frac{R^2 A_{11} + RB_{11}}{A_{11}D_{11} - B_{11}^2} + \frac{RB_{11} + D_{11}}{A_{11}D_{11} - B_{11}^2} \right) \tag{14}$$

And the constants $\lambda_1$ and $\lambda_2$ in the first equation of Eq. (13) are given as

$$\lambda_1 = 2 - \frac{k_r R^2}{A_{55}}, \lambda_2 = 1 + \frac{k_r R^2 (R^2 A_{11} + RB_{11})}{A_{11}D_{11} - B_{11}^2} + \frac{k_r R^2 (RB_{11} + D_{11})}{A_{11}D_{11} - B_{11}^2} \tag{15}$$

Since the values of $\lambda_1$ and $\lambda_2$ determine the different forms of general solutions of Eq. (13), we discuss two cases of relationships between $\lambda_1$ and $\lambda_2$.

**Case 1:**

When $\lambda_1^2 - 4\lambda_2 \geq 0$, which gives

$$k_r \geq \frac{4A_{55}}{R^2} \left[ 1 + \frac{A_{55}(R^2 A_{11} + RB_{11})}{A_{11}D_{11} - B_{11}^2} + \frac{A_{55}(RB_{11} + D_{11})}{A_{11}D_{11} - B_{11}^2} \right] \tag{16}$$

the general solutions of Eq. (13) are solved as

$$\begin{aligned}
&u_r(\theta) = C_1 \sinh \alpha_1 \theta + C_2 \cosh \alpha_1 \theta + C_3 \sinh \beta_1 \theta + C_4 \cosh \beta_1 \theta + C_5 \\
&u_{\theta 0}(\theta) = C_1 Q_1 \cosh \alpha_1 \theta + C_2 Q_1 \sinh \alpha_1 \theta + C_3 Q_2 \cosh \beta_1 \theta + C_4 Q_2 \sinh \beta_1 \theta - \\
&C_5 \frac{k_r R^3 + RA_{11}}{RA_{11} + B_{11}} \theta + C_6 \\
&\phi(\theta) = C_1 H_1 \cosh \alpha_1 \theta + C_2 H_1 \sinh \alpha_1 \theta + C_3 H_2 \cosh \beta_1 \theta + C_4 H_2 \sinh \beta_1 \theta \\
&-C_5 \frac{k_r R^2 + A_{11}}{RA_{11} + B_{11}} \theta + \frac{C_6}{R}
\end{aligned} \tag{17}$$

where the constants ($\alpha_1$, $\beta_1$, $Q_1$, $Q_2$, $H_1$, $H_2$) used in Eq. (17) are defined as

$$\begin{aligned}
\alpha_1 &= \sqrt{-\frac{1}{2}\lambda_1 + \sqrt{\frac{1}{4}\lambda_1^2 - \lambda_2}} \\
\beta_1 &= \sqrt{-\frac{1}{2}\lambda_1 - \sqrt{\frac{1}{4}\lambda_1^2 - \lambda_2}} \\
Q_1 &= \frac{\alpha_1^3 + B_{11}\alpha_1 P + \alpha_1}{P(RA_{11} + B_{11})} - \frac{\alpha_1 k_r R^2}{PA_{55}(RA_{11} + B_{11})} - \frac{k_r R^3 + RA_{11}}{\alpha_1(RA_{11} + B_{11})} \\
Q_2 &= \frac{\beta_1^3 + B_{11}\beta_1 P + \beta_1}{P(RA_{11} + B_{11})} - \frac{\beta_1 k_r R^2}{PA_{55}(RA_{11} + B_{11})} - \frac{k_r R^3 + RA_{11}}{\beta_1(RA_{11} + B_{11})} \\
H_1 &= \frac{Q_1 - \alpha_1}{R} - \frac{(\alpha_1^2 Q_1 + \alpha_1)(A_{11}D_{11} - B_{11}^2)}{RA_{55}(RB_{11} + D_{11})} \\
H_2 &= \frac{Q_2 - \beta_1}{R} - \frac{(\beta_1^2 Q_2 + \beta_1)(A_{11}D_{11} - B_{11}^2)}{RA_{55}(RB_{11} + D_{11})}
\end{aligned} \tag{18}$$

Considering the symmetry at $\theta = 0$, one can provide boundary conditions on the top of the ring as

$$u_{\theta 0}(0) = 0, \phi(0) = 0, V(0) = 0 \tag{19}$$

Note that Eq. (19) can eliminate three unknown constants ($C_1$, $C_3$, and $C_6$) in Eq. (17). Then, considering the symmetry at $\theta = \pi$, we get boundary conditions on the bottom of the ring as

$$u_{\theta 0}(\pi) = 0, \phi(\pi) = 0, V(\pi) = -F/2 \tag{20}$$

Therefore, the remaining three unknown constants ($C_2$, $C_4$, and $C_5$) in Eq. (17) are solved as follows:

$$\begin{aligned}
C_2 &= -\frac{FR(Q_2 - RH_2)}{2A_{55}(\alpha_1 Q_2 - \beta_1 Q_1 - R\alpha_1 H_2 + R\beta_1 H_1)\sinh\alpha_1\pi} \\
C_4 &= \frac{FR(Q_1 - RH_1)}{2A_{55}(\alpha_1 Q_2 - \beta_1 Q_1 - R\alpha_1 H_2 + R\beta_1 H_1)\sinh\beta_1\pi} \\
C_5 &= \frac{FR(B_{11} + RA_{11})(Q_1 H_2 - Q_2 H_1)}{2\pi A_{55}(k_r R^2 + A_{11})(\alpha_1 Q_2 - \beta_1 Q_1 - R\alpha_1 H_2 + R\beta_1 H_1)}
\end{aligned} \tag{21}$$

**Case 2:**

When $\lambda_1^2 - 4\lambda_2 < 0$, which gives

$$k_r < \frac{4A_{55}}{R^2}\left[1 + \frac{A_{55}(R^2 A_{11} + RB_{11})}{A_{11}D_{11} - B_{11}^2} + \frac{A_{55}(RB_{11} + D_{11})}{A_{11}D_{11} - B_{11}^2}\right] \tag{22}$$

the general solutions of Eq. (13) become

$$u_r(\theta) = C_1 \sinh \alpha_2\theta \cos \beta_2\theta + C_2 \cosh \alpha_2\theta \cos \beta_2\theta + C_3 \cosh \alpha_2\theta \sin \beta_2\theta + C_4 \sinh \alpha_2\theta \sin \beta_2\theta + C_5$$
$$u_{\theta 0}(\theta) = (C_1\alpha_2 Q_3 + C_3\beta_2 Q_4) \cosh \alpha_2\theta \cos \beta_2\theta + (C_2\alpha_2 Q_3 + C_4\beta_2 Q_4) \sinh \alpha_2\theta \cos \beta_2\theta + (C_3\alpha_2 Q_3 - C_1\beta_2 Q_4) \sinh \alpha_2\theta \sin \beta_2\theta + (C_4\alpha_2 Q_3 - C_2\beta_2 Q_4) \cosh \alpha_2\theta \sin \beta_2\theta - C_5 \frac{k_r R^3 + RA_{11}}{RA_{11} + B_{11}}\theta + C_6 \quad (23)$$
$$\phi(\theta) = (C_1\alpha_2 H_3 - C_3\beta_2 H_4) \cosh \alpha_2\theta \cos \beta_2\theta + (C_2\alpha_2 H_3 - C_4\beta_2 H_4) \sinh \alpha_2\theta \cos \beta_2\theta + (C_3\alpha_2 H_3 + C_1\beta_2 H_4) \sinh \alpha_2\theta \sin \beta_2\theta + (C_4\alpha_2 H_3 + C_2\beta_2 H_4) \cosh \alpha_2\theta \sin \beta_2\theta - C_5 \frac{k_r R^2 + A_{11}}{RA_{11} + B_{11}}\theta + \frac{C_6}{R}$$

Here, the constants ($\alpha_2$, $\beta_2$, $Q_3$, $Q_4$, $H_3$, $H_4$) in Eq. (23) are defined as

$$\alpha_2 = \sqrt{\frac{1}{2}\sqrt{\lambda_2} - \frac{1}{4}\lambda_1}$$
$$\beta_2 = \sqrt{\frac{1}{2}\sqrt{\lambda_2} + \frac{1}{4}\lambda_1}$$
$$Q_3 = \frac{\alpha_2^2 - 3\beta_2^2 + B_{11}P + 1}{P(RA_{11} + B_{11})} - \frac{k_r R^2}{PA_{55}(RA_{11} + B_{11})} - \frac{k_r R^3 + RA_{11}}{(\alpha_2^2 + \beta_2^2)(RA_{11} + B_{11})} \quad (24)$$
$$Q_4 = \frac{3\alpha_2^2 - \beta_2^2 + B_{11}P + 1}{P(RA_{11} + B_{11})} - \frac{k_r R^2}{PA_{55}(RA_{11} + B_{11})} + \frac{k_r R^3 + RA_{11}}{(\alpha_2^2 + \beta_2^2)(RA_{11} + B_{11})}$$
$$H_3 = \frac{[2\beta_2^2 Q_4 - (\alpha_2^2 - \beta_2^2)Q_3 - 1](A_{11}D_{11} - B_{11}^2)}{RA_{55}(RB_{11} + D_{11})} + \frac{Q_3 - 1}{R}$$
$$H_4 = \frac{[2\alpha_2^2 Q_3 + (\alpha_2^2 - \beta_2^2)Q_4 + 1](A_{11}D_{11} - B_{11}^2)}{RA_{55}(RB_{11} + D_{11})} - \frac{Q_4 - 1}{R}$$

Similarly, the boundary conditions on the top of the ring with a symmetry at $\theta = 0$ provides

$$u_{\theta 0}(0) = 0, \phi(0) = 0, V(0) = 0 \quad (25)$$

based on which $C_1$, $C_3$, and $C_6$ are canceled in Eq. (23). Besides, the boundary conditions on the bottom of the ring with symmetry at $\theta = \pi$ provides

$$u_{\theta 0}(\pi) = 0, \phi(\pi) = 0, V(\pi) = -F/2 \quad (26)$$

Therefore, $C_2$, $C_4$, and $C_5$ in Eq. (23) are solved as follows:

$$C_2 = \frac{FR(\alpha_2 Q_3 \cosh \alpha_2\pi \sin \beta_2\pi + \beta_2 Q_4 \sinh \alpha_2\pi \cos \beta_2\pi)}{2A_{55}\alpha_2\beta_2(Q_3 - Q_4 - RH_3 - RH_4)(\cosh^2 \alpha_2\pi \sin^2 \beta_2\pi + \sinh^2 \alpha_2\pi \cos^2 \beta_2\pi)} - \frac{FR^2(\alpha_2 H_3 \cosh \alpha_2\pi \sin \beta_2\pi - \beta_2 H_4 \sinh \alpha_2\pi \cos \beta_2\pi)}{2A_{55}\alpha_2\beta_2(Q_3 - Q_4 - RH_3 - RH_4)(\cosh^2 \alpha_2\pi \sin^2 \beta_2\pi + \sinh^2 \alpha_2\pi \cos^2 \beta_2\pi)}$$
$$C_4 = -\frac{FR(\alpha_2 Q_3 \sinh \alpha_2\pi \cos \beta_2\pi - \beta_2 Q_4 \cosh \alpha_2\pi \sin \beta_2\pi)}{2A_{55}\alpha_2\beta_2(Q_3 - Q_4 - RH_3 - RH_4)(\cosh^2 \alpha_2\pi \sin^2 \beta_2\pi + \sinh^2 \alpha_2\pi \cos^2 \beta_2\pi)} + \frac{FR^2(\alpha_2 H_3 \sinh \alpha_2\pi \cos \beta_2\pi + \beta_2 H_4 \cosh \alpha_2\pi \sin \beta_2\pi)}{2A_{55}\alpha_2\beta_2(Q_3 - Q_4 - RH_3 - RH_4)(\cosh^2 \alpha_2\pi \sin^2 \beta_2\pi + \sinh^2 \alpha_2\pi \cos^2 \beta_2\pi)} \quad (27)$$
$$C_5 = \frac{FR(RA_{11} + B_{11})(Q_3 H_4 + Q_4 H_3)}{2\pi A_{55}(k_r R^2 + A_{11})(Q_3 - Q_4 - RH_3 - RH_4)}$$

In Eqs. (21) and (27), all constants $C_i$ are linearly related to the point load $F$. Thus, the normalized radial displacement $\bar{u}_r(\theta)$, circumferential displacement $\bar{u}_{\theta 0}(\theta)$, and in-plane cross-section's rotation

$\bar{\phi}(\theta)$ can be expressed as

$$\begin{aligned} \bar{u}_r(\theta) &= u_r(\theta)/F \\ \bar{u}_{\theta 0}(\theta) &= u_{\theta 0}(\theta)/F \\ \bar{\phi}(\theta) &= \phi(\theta)/F \end{aligned} \tag{28}$$

3.2. A laminated ring on a linear elastic foundation subjected to a radially distributed load

The vertical point loading scenario on the bottom of the ring in Section 3.1 can be extended to a radial point loading at any circumferential position, as illustrated in Figure 2b. For a vertical point load, one can rotate the original coordinate in Figure 2a by $\theta_F$, defining a new coordinate with the $X'Y'$ coordinate. The position in the new coordinate is described with $\theta'$, where $\theta' = \theta - \theta_F$. Here, $\theta_F > 0$ signifies a clockwise rotation of $F$ and $\theta_F < 0$ implies a counterclockwise rotation of $F$. Note that the position of the load $\theta_F$ is measured from the bottom of the ring. Consequently, the general case of the normalized displacements in Eq.(28) is modified as

$$\begin{aligned} \bar{u}_r(\theta, \theta_F) &= \bar{u}_r(\theta - \theta_F) \\ \bar{u}_{\theta 0}(\theta, \theta_F) &= \bar{u}_{\theta 0}(\theta - \theta_F) \\ \bar{\phi}(\theta, \theta_F) &= \bar{\phi}(\theta - \theta_F) \end{aligned} \tag{29}$$

where $\bar{u}_r(\theta, \theta_F)$, $\bar{u}_{\theta 0}(\theta, \theta_F)$, and $\bar{\phi}(\theta, \theta_F)$ are the normalized radial displacement, circumferential displacement, and in-plane cross-sectional rotation at $\theta$ produced by a radial point load at the position $\theta_F$.

Then, one can extend the deformation of a ring with a point loading concept to formulate the deformation with a distributed load. A point load $dF(\theta_F)$ applied on an infinitesimal element in Figure 2c can be expressed as

$$dF(\theta_F) = w(\theta_F) ds_F \tag{30}$$

where $dF(\theta_F)$ denotes the point load applied on the infinitesimal element at the position $\theta_F$, $w(\theta_F)$ is an equivalent distributed load on the element at $\theta_F$. The element's length $ds_F$ on the exterior edge of the ring is given by $ds_F = (R + h/2) d\theta_F$. Therefore, the displacements and rotation of the ring under a distributed load are given via the superposition of solutions for each point load, expressed by integral forms within the range [$\theta_1$, $\theta_2$]

$$\begin{aligned} U_r(\theta) &= \int_{\theta_1}^{\theta_2} dF(\theta_F) \bar{u}_r(\theta, \theta_F) = \int_{\theta_1}^{\theta_2} w(\theta_F) \bar{u}_r(\theta, \theta_F)(R + h/2)\, d\theta_F \\ U_{\theta 0}(\theta) &= \int_{\theta_1}^{\theta_2} dF(\theta_F) \bar{u}_{\theta 0}(\theta, \theta_F) = \int_{\theta_1}^{\theta_2} w(\theta_F) \bar{u}_{\theta 0}(\theta, \theta_F)(R + h/2)\, d\theta_F \\ \Phi(\theta) &= \int_{\theta_1}^{\theta_2} dF(\theta_F) \bar{\phi}(\theta, \theta_F) = \int_{\theta_1}^{\theta_2} w(\theta_F) \bar{\phi}(\theta, \theta_F)(R + h/2)\, d\theta_F \end{aligned} \tag{31}$$

where $U_r(\theta)$, $U_{\theta 0}(\theta)$, and $\Phi(\theta)$ are the radial displacement, circumferential displacement, and cross-section rotation on the centroid of a ring subjected to a distributed load. Rewriting Eq. (31) in terms of the summation notation yields

$$
\begin{aligned}
U_r(\theta) &= \sum_{i=1}^{N_F} w(\theta_F)\bar{u}_r(\theta,\theta_F)(R+h/2)\Delta\theta_F \\
U_{\theta 0}(\theta) &= \sum_{i=1}^{N_F} w(\theta_F)\bar{u}_{\theta 0}(\theta,\theta_F)(R+h/2)\Delta\theta_F \\
\Phi(\theta) &= \sum_{i=1}^{N_F} w(\theta_F)\bar{\phi}(\theta,\theta_F)(R+h/2)\Delta\theta_F
\end{aligned}
\tag{32}
$$

where $N_F$ is the number of elements divided along the load region. As we equally divide the loaded region into $N_F$ elements in this context, each element has an angle of $\Delta\theta_F$.

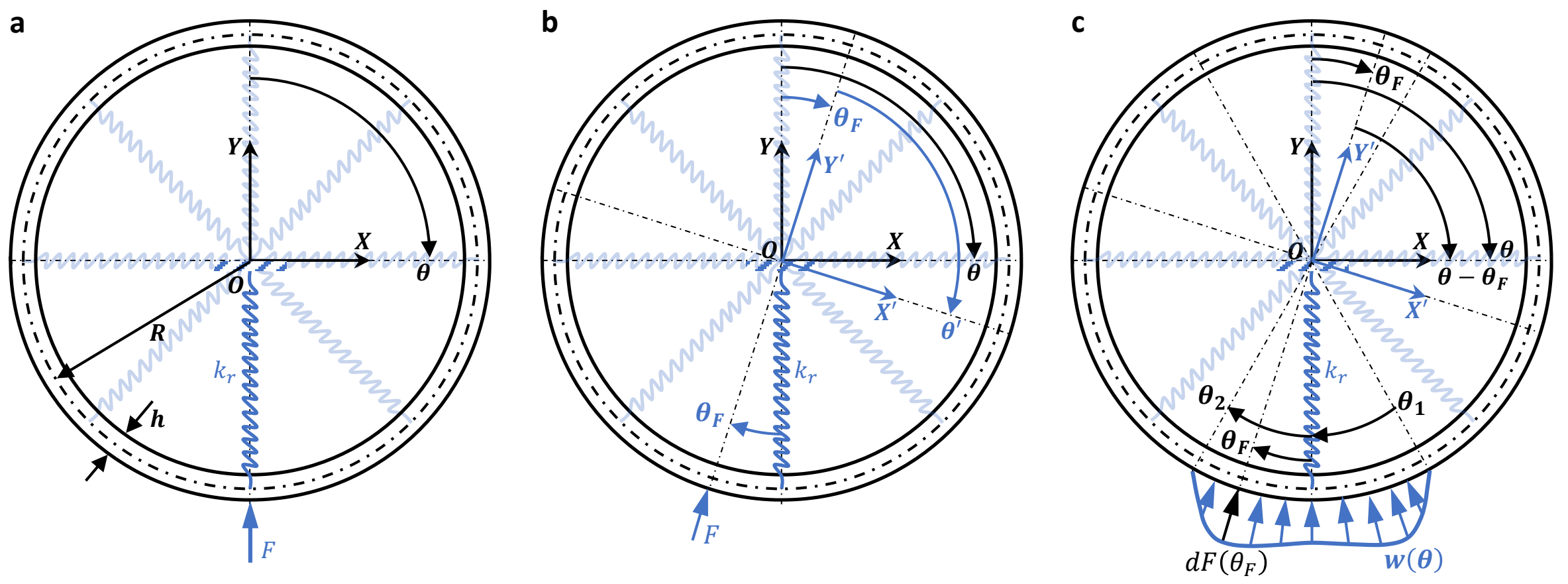

*Figure 2. Construction of a laminated ring on a linear foundation (LRLF) model: (a) a laminated ring subjected to a radial point load at $\theta = \pi$; (b) a laminated ring subjected to a radial point load applied at an arbitrary position $\theta_F$; (c) a laminated ring subjected to a distributed load.*

### 3.3. A laminated ring on a nonreciprocal elastic foundation (LRNF) subjected to a radial point load

The displacement solutions in the previous sections are derived for rings on a linear foundation, where the foundation's stiffness $k_r$ is kept constant whether the foundation is stretched or compressed. In some engineering applications, such as collapsible spokes in non-pneumatic wheels [36], the equivalent elastic foundation exhibits an asymmetric transmission of mechanical loading for compression and tension, called elastic nonreciprocity, requiring an advanced elastic foundation model.

As shown in Figure 3a, assuming the nonreciprocal foundation, we set different radial stiffness on the elastic foundation for tensile and compressive modes; $k_{rT} \neq k_{rC}$, where $k_{rT}$ is the tensile stiffness and $k_{rC}$ is the compressive stiffness of the elastic foundation. Note that we only consider the case of $k_{rT} > k_{rC}$, since the derivatives of the solution for $k_{rT} < k_{rC}$ are similar. Notably, due to the smaller compression stiffness in the nonreciprocal foundation, a larger deformation can be generated in the compressive region (where $u_r(\theta) < 0$) compared with that in the linear foundation with $k_{rT} = k_{rC}$.

To obtain accurate deformation for the nonreciprocal foundation case, we apply a compensation load $w_{cp}^{(i)}(\theta)$ to the ring on a linear foundation, eliminating the displacement errors due to the stiffness difference between tension and compression. The amount of this compensation load $w_{cp}^{(i)}$ is proportional to the negative radial displacement in the initial solution of the linear foundation case, which is given by

$$w_{cp}^{(i)}(\theta) = \begin{cases} (k_{rT} - k_{rC})u_r^{(i)}(\theta), & u_r^{(i)}(\theta) < 0 \\ 0, & u_r^{(i)}(\theta) \geq 0 \end{cases} \tag{33}$$

Here, the superscript $i$ denotes the index for the current iteration step. For $i = 0$, $u_r^{(0)}$ is the radial displacement of the laminated ring equipped with a linear foundation with $k_{rT}$ ($= k_{rC}$). Then, the compensation load is applied to the exterior edge of the ring. Utilizing the method proposed in Section 3.2, we get the compensation displacements $u_{r-cp}^{(i)}(\theta)$, $u_{\theta 0-cp}^{(i)}(\theta)$, and $\phi_{cp}^{(i)}(\theta)$. Adding these compensation displacements to the original solutions gives the initial solution for the nonreciprocal foundation case, which are

$$\begin{aligned} u_r^{(i+1)}(\theta) &= u_r^{(i)}(\theta) + u_{r-cp}^{(i)}(\theta) \\ u_{\theta 0}^{(i+1)}(\theta) &= u_{\theta 0}^{(i)}(\theta) + u_{\theta 0-cp}^{(i)}(\theta) \\ \phi^{(i+1)}(\theta) &= \phi^{(i)}(\theta) + \phi_{cp}^{(i)}(\theta) \end{aligned} \tag{34}$$

The update of radial displacement from $u_r^{(i)}(\theta)$ to $u_r^{(i+1)}(\theta)$ results in a new compensation load $w_{cp}^{(i+1)}(\theta)$, which generates new compensation displacements $u_{r-cp}^{(i+1)}$. Figure 3b illustrates the flowchart of the numerical scheme to obtain the deformation of a laminated ring equipped with a nonreciprocal foundation. The algorithm's iterative procedure from Eq. (33) to Eq. (34) must be repeated until the compensation displacements converge. In such a case, the summation of displacements in Eq. (34) is the final solution for the nonreciprocal foundation case.

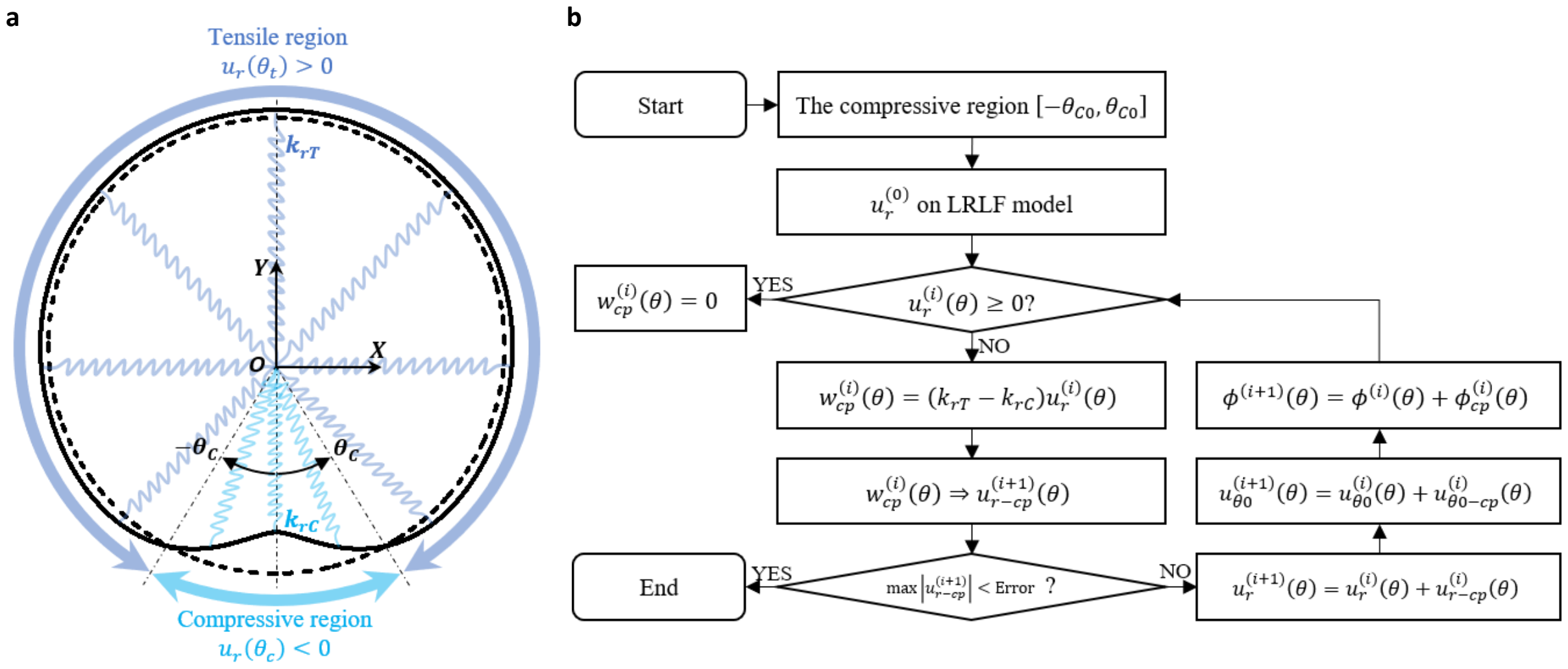

*Figure 3. Construction of a laminated ring on a nonreciprocal foundation (LRNF) model: (a) illustration of a loaded laminated ring subjected to a radial point load; (b) the compensation algorithm for obtaining the deformation of the laminated ring on a nonreciprocal foundation (detailed in Appendix B).*

## 3.4. Validation of model by FE simulations

To validate the analytical results in Sections 3.1 and 3.3, we simulate the ring model with finite element software ABAQUS/Standard. The validation process encompasses a composite material with a lay-up defined as [90°/0°/90°/0°/90°]. We discretize the thick ring depicted in Figure 2 using the 4-node bilinear quadrilateral plane strain element with reduced integration (CPE4R in ABAQUS/Standard), with a 10-layer division in the radial direction and 480 elements divided in the circumferential direction of the ring. To implement the elastic foundation in Figure 2, we discretize the elastic foundation with 120 spring elements (SpringA in ABAQUS/Standard), evenly distributed in the circumferential direction. Thus, the spring stiffness $k_{sp}$ is calculated as

$$k_{sp} = \frac{2\pi R k_r}{N_{sp}} \tag{35}$$

where $N_{sp}$ denotes the number of spring elements. To implement the nonreciprocal elastic foundation, we set the radial tension and compression stiffnesses of the elastic foundation as $k_{rT} = 1\ \mathrm{N/mm^2}$ and $k_{rC} = 0.1\ \mathrm{N/mm^2}$ in ABAQUS, equivalent to $k_{spT} = 10.47\ \mathrm{N/mm}$ and $k_{spC} = 1.047\ \mathrm{N/mm}$ in Eq. (35). The geometric parameters of the ring are specified as $R = 200\ \mathrm{mm}$, $h = 20\ \mathrm{mm}$, and $b = 60\ \mathrm{mm}$. The mechanical properties of the lamina are $E_{11} = 20.38\ \mathrm{GPa}$, $E_{22} = E_{33} = 13.06\ \mathrm{MPa}$, $G_{12} = G_{13} = 3.50\ \mathrm{MPa}$, $G_{23} = 3.43\ \mathrm{MPa}$, $\nu_{12} = \nu_{13} = 0.39$, $\nu_{23} = 0.89$, obtained based on the Halpin-Tsai model [42]. From the analytical model, the deformed ring's exterior positions along the radial and circumferential directions are calculated as

$$\begin{aligned} \theta_e(R + h/2\,, \theta) &= \theta + \tan^{-1}\frac{u_\theta}{R + h/2 + u_r} \\ r_e(R + h/2\,, \theta) &= \sqrt{(R + h/2 + u_r)^2 + u_\theta^2} \end{aligned} \tag{36}$$

Figure 4a compares the deformed configuration of the ring's exterior edge obtained by the analytical model and FE simulations for linear and nonreciprocal foundation cases. The magnitude of the point load is $F = 3\ \mathrm{kN}$. This comparison demonstrates a remarkable match of the deformation profiles. Furthermore, Figure 4b and Figure 4c illustrate that the radial and circumferential displacements along the middle surface of the laminated ring calculated by the analytical model align closely with the results from the FE simulations. Notably, the nonreciprocal foundation produces a more significant radial displacement over the ring's top region and a more significant compressive displacement over the ring's bottom region compared to the linear elastic foundation, as shown in Figure 4b. The variation of circumferential displacement of the ring is greater with the nonreciprocal foundation compared with the linear counterpart, as shown in Figure 4c.

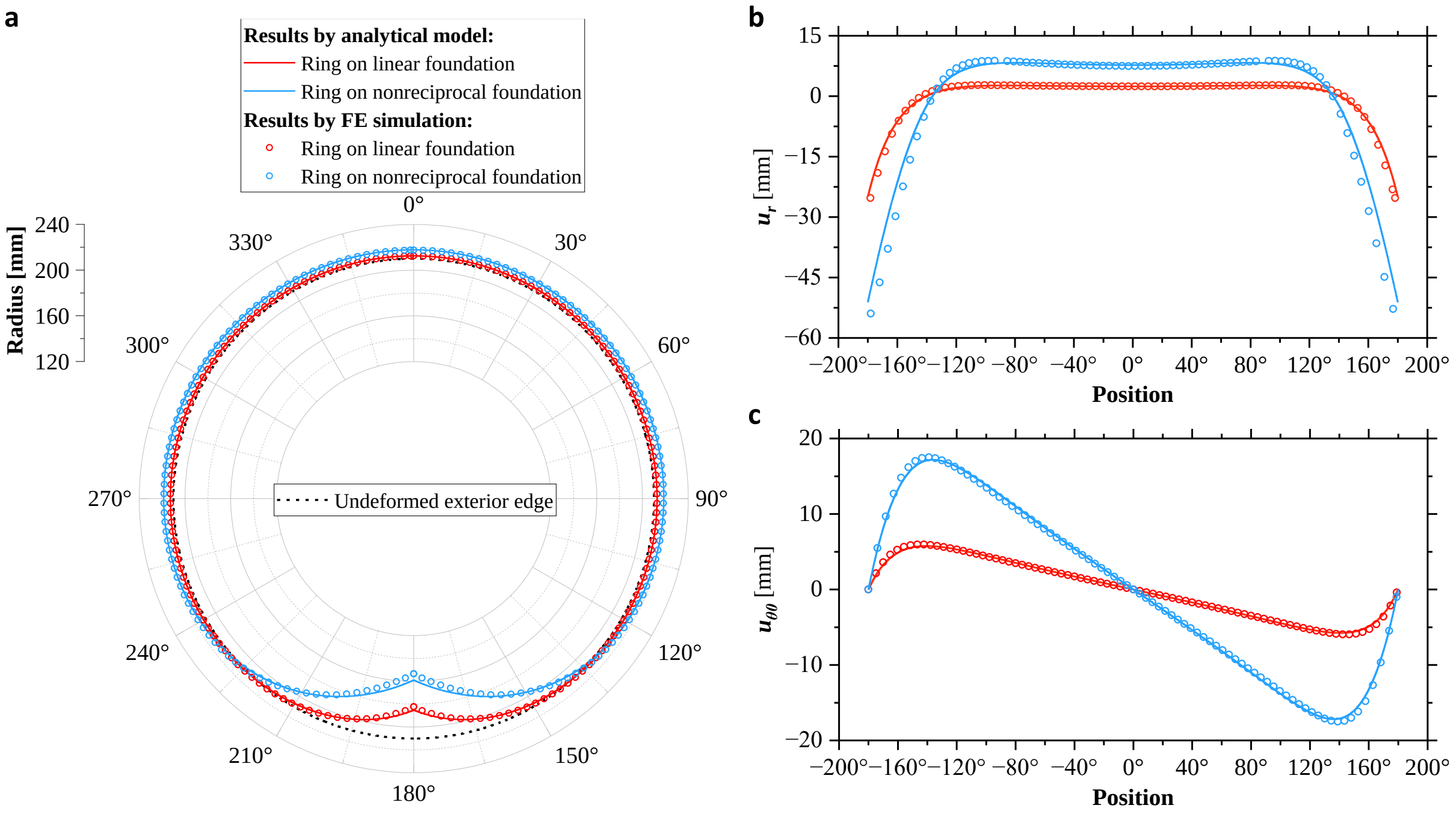

*Figure 4. Validation of the laminated ring model by FE simulations for linear and nonreciprocal foundations: (a) deformed configurations of the ring's exterior edge under the vertical point load at $\theta = \pi$; (b) and (c) are the radial and circumferential displacements along the middle surface of the laminated ring, respectively.*

## 4. Static analysis of the laminated ring on an elastic foundation

### 4.1. A ring system with a linear elastic foundation

For a static analysis of a laminated ring interacting with an elastic foundation, we quantify $\bar{K}_{LF}$, the ratio of the overall stiffness $K_{LF}$ of the ring-foundation system to the laminated ring stiffness $K$.

$$\bar{K}_{LF} = \frac{K_{LF}}{K} = \frac{1}{K}\frac{F}{\delta} = \frac{1}{K|\bar{u}_r(\pi)|} \tag{37}$$

Notably, a laminated ring's stiffness $K$ is a function of geometry and material properties, as expressed by

$$K = \frac{8\pi}{\pi^2 \dfrac{R}{GA} + \pi^2\left(\dfrac{R}{EA} + \dfrac{R^2}{EV}\right) + \left(\pi^2 - 8 - \dfrac{8EI}{REV}\right)\left(\dfrac{R^3}{EI} + \dfrac{R^2}{EV}\right)} \tag{38}$$

A detailed derivation procedure of Eq. (38) is shown in Appendix A. The ratio $\bar{K}_{LF}$ identifies the deformation mode of the ring system – a laminated ring on the elastic foundation. For instance, $\bar{K}_{LF} \leq 1$ implies that the overall deformation is mainly affected by the elastic foundation due to the greater ring's stiffness $K$. On the other hand, $\bar{K}_{LF} > 1$ means a smaller ring's stiffness $K$ dominantly affects the overall deformation of the ring system. The effective stiffness parameters ($EA$, $EV$, $EI$, and $GA$) used in Eq. (38) are obtained by inverting the compliance equations in Eq. (7):

$$\begin{bmatrix} \varepsilon_{\theta\theta}^0 \\ \kappa \\ \gamma_{r\theta}^0 \end{bmatrix} = \begin{bmatrix} 1/EA & -1/EV & 0 \\ -1/EV & 1/EI & 0 \\ 0 & 0 & 1/GA \end{bmatrix} \begin{bmatrix} N \\ M \\ V \end{bmatrix} \tag{39}$$

where $EA$ is the effective extension stiffness, $EI$ is the effective bending stiffness, $EV$ is the effective bending-extension coupling stiffness, and $GA$ is the effective shear stiffness of a laminate, given by

$$EA = A_{11} - \frac{B_{11}^2}{D_{11}}, EV = \frac{A_{11}D_{11}}{B_{11}} - B_{11}, EI = D_{11} - \frac{B_{11}^2}{A_{11}}, GA = A_{55} \tag{40}$$

Figure 5a presents the normalized load-deflection ($\bar{F} - \bar{\delta}$) curves for a laminated ring on a linear foundation for varying $\bar{K}_{LR}$. Here, normalized deflection ($\bar{\delta}$) is defined as $\bar{\delta} = \delta/R$, and normalized load ($\bar{F}$) is defined by $\bar{F} = F/(RK)$. For all cases in Figure 5a, the orthotropic material properties are $E_2 = 20$ MPa, $E_1/E_2 = 1000$, $G_{12}/E_2 = 0.5$, $G_{23}/E_2 = 0.3$, $G_{13} = G_{12}$, and $\nu_{12} = 0.4$. The midplane radius of the four-layered $[0°, 90°]_s$ ring is $R = 200$ mm with an overall thickness of $h = 20$ mm and an out-of-plane width of $b = 60$ mm. A stiffer elastic foundation, e.g., $\bar{K}_{LF} = 5.0$, dominantly influences the overall stiffness of the ring system, as shown in Figure 5a. The stiffer elastic foundation leads to more pronounced local deformation for a point load compared to the system with a stiffer ring ($\bar{K}_{LF} = 0.5$) under the identical vertical deflection ($\bar{\delta} = 0.15$), as shown in Figure 5b. Notably, our analytical model agrees well with FE simulations of ABAQUS, where a nonlinear geometry solver is used.

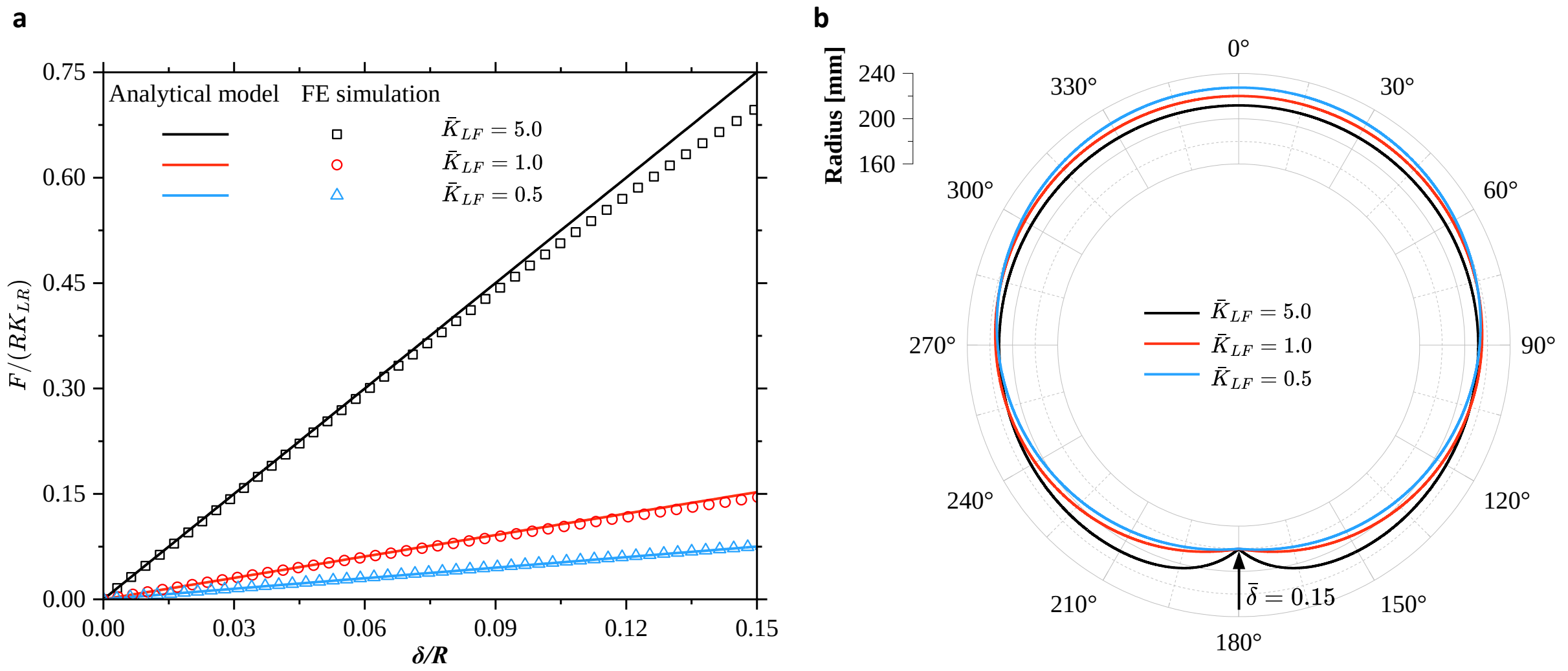

*Figure 5. Static behavior of a laminated ring system for varying $\bar{K}_{LF}$: (a) the normalized load-deflection ($\bar{F} - \bar{\delta}$) curves; (b) deformations of the exterior edge of the ring under a vertical point load.*

Figure 6 shows a parametric analysis to investigate the effects of the thickness ratio ($h/R$) and foundation modulus ($k_r$) on $\bar{K}_{LF}$ for varying orthotropy ratios ($E_1/E_2 = 1000, 100, 10,$ and $1$). The material properties and layer stacking of laminated rings are consistent with those for the analysis in Figure 5. The solid black lines with arrows in Figure 6 indicate the stiffness of laminated rings ($K$) as a function of $h/R$, showing $K$ increases with $h/R$. Figure 6 shows that as the orthotropy ratio ($E_1/E_2$) decreases, there is a corresponding increase in $\bar{K}_{LF}$, assuming $E_2$ remains constant. An increase in $h$ for a constant $R$ makes the ring's stiffness $K$ greater in Eq. (38), decreasing $\bar{K}_{LF}$ in Eq. (37), as confirmed in Figure 6. Figure 6 also confirms that a higher foundation modulus $k_r$ makes $K_{LF}$ greater, increasing $\bar{K}_{LF}$ in Eq. (37). A higher $\bar{K}_{LF}$ ($\bar{K}_{LF} > 1$) indicates a stiffer elastic foundation and a weaker ring, resulting in significant local deformation of the ring under an external point load. Conversely, a lower $\bar{K}_{LF}$ ($\bar{K}_{LF} \leq 1$) suggests the foundation is softer relative to the ring, minimizing local ring deformation, as shown in Figure 6. The dotted lines of $\bar{K}_{LF} = 1$ can serve as a design criterion for the ring system to avoid local deformation.

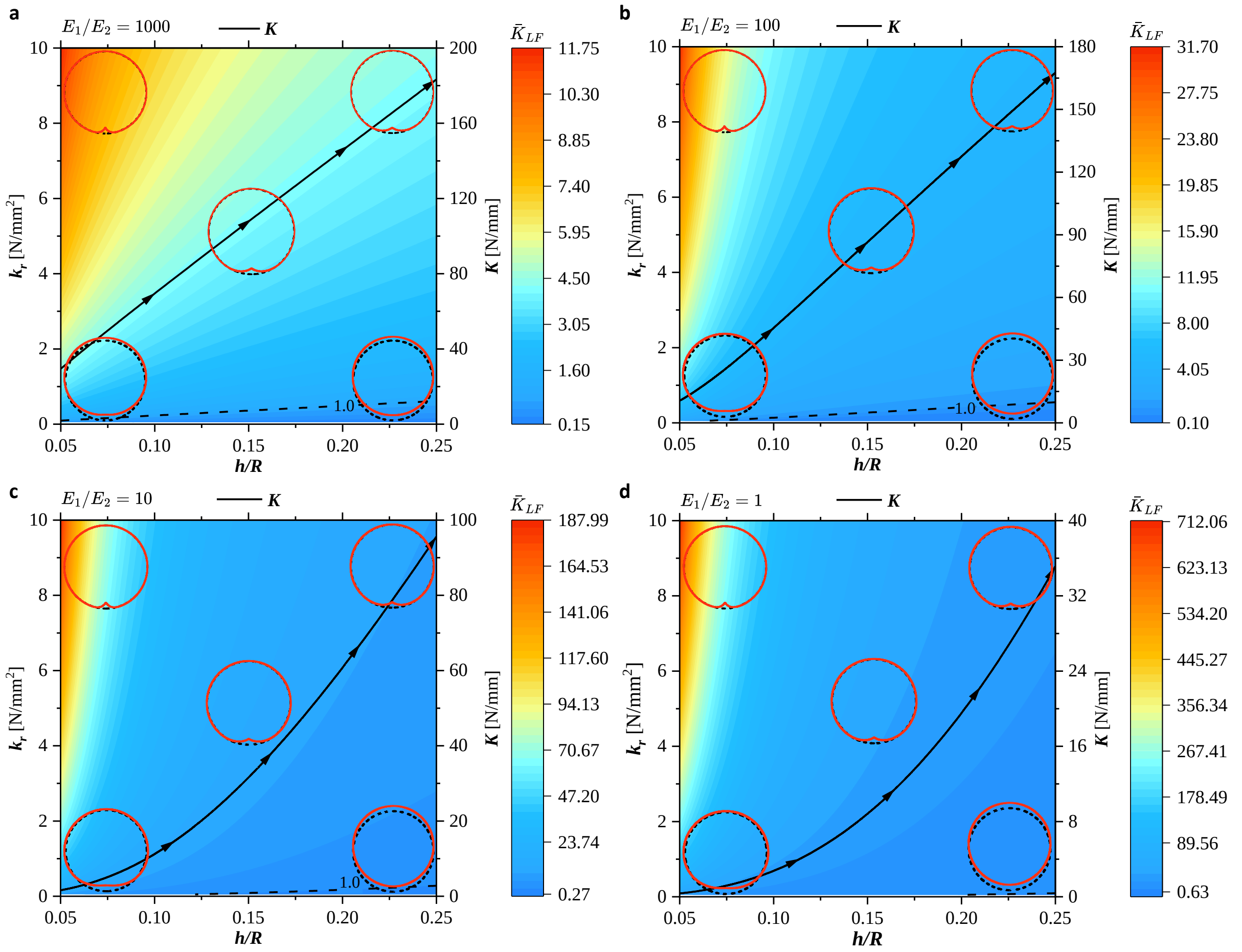

*Figure 6. Sensitivity of $\bar{K}_{LF}$ to the thickness ratio $h/R$ and the foundation modulus $k_r$ for varying orthotropy ratios $E_1/E_2$: (a) $E_1/E_2 = 1{,}000$, (b) $E_1/E_2 = 100$, (c) $E_1/E_2 = 10$, and (d) $E_1/E_2 = 1$.*

## 4.2. A ring system with a nonreciprocal elastic foundation

Figure 7a shows the normalized load-deflection ($\bar{F} - \bar{\delta}$) curves of the laminated ring on a nonreciprocal foundation (NF) for varying nonreciprocal ratios ($k_{rC}/k_{rT}$). Note that $k_{rC}/k_{rT} = 1$ corresponds to the linear foundation ($k_{rC} = k_{rT} = k_r$) and $k_{rC}/k_{rT} = 0$ corresponds to a unilateral foundation ($k_{rC} = 0$) [13]. For the same ring stiffness $K$, the system becomes weaker with a lower $k_{rC}/k_{rT}$, as confirmed by the analytical models and FE simulations in Figure 7. Figure 7b illustrates the deformation responses of the rings' exterior surface under a point load of $F = 2{,}400\text{N}$ for interacting with the nonreciprocal elastic foundation. The figure demonstrates that the magnitude of radial displacement $u_r$ increases as $k_{rC}/k_{rT}$ decreases at the loading region on the bottom. To balance the external point load $F$, the tensile region compensates by generating a larger radial displacement to offset the reduced load-bearing capacity in the compressive region. The diminished resistance in the compressive region due to the nonreciprocal elastic foundation induces more significant local deformation than the linear foundation case.

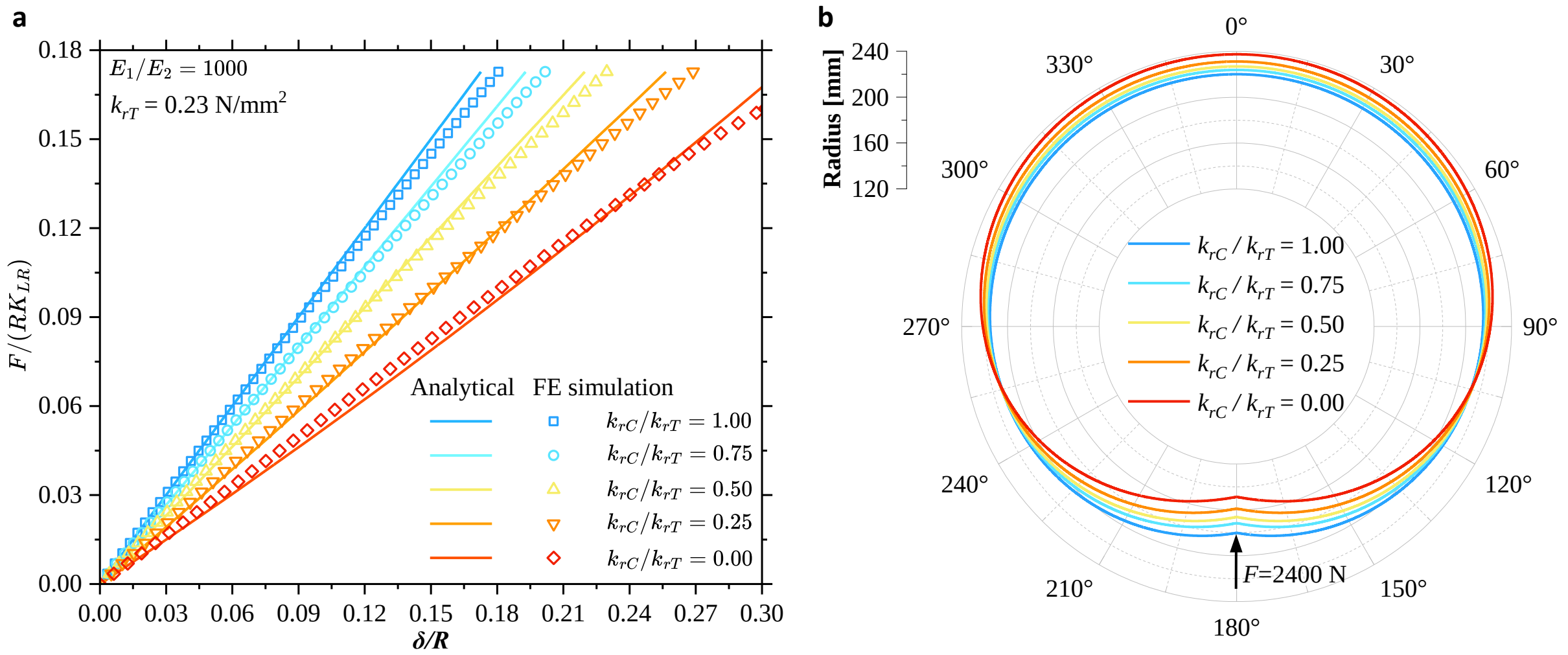

*Figure 7. Static behavior of a laminated ring on nonreciprocal elastic foundations for varying nonreciprocal ratios $k_{rC}/k_{rT}$: (a) the normalized load-deflection ($\bar{F} - \bar{\delta}$) curves; (b) deformation responses of the rings' exterior surface under a vertical point load of $F = 2{,}400\ N$.*

One can extend the definition of Eq. (37) by defining $\overline{K}_{NF}$, the ratio of the nonreciprocal elastic foundation $K_{NF}$ to the laminated ring stiffness $K$; $\overline{K}_{NF} = K_{NF}/K$. Figure 8a shows $\overline{K}_{NF}$ for varying $h/R$ and $k_r$ where we consider two nonreciprocal ratios in the elastic foundation, where $k_{rC}/k_{rT} = 0$ and $k_{rC}/k_{rT} = 0.5$. Note that $k_{rT} = k_r$. The orthotropy ratio is selected as $E_1/E_2 = 1{,}000$. The material properties and layer stacking of the laminated ring are consistent with those shown in Figure 5. Compared to the ring system of a linear elastic foundation in Figure 6a, the ring system with a nonreciprocal elastic foundation produces a lower $\overline{K}_{NF}$, implying a lower foundation stiffness $K_{NF}$, as shown in Figure 8a and Figure 8b. Notably, the dotted line of $\overline{K}_{NF} = 1$ in Figure 8a and Figure 8b shifts upward compared to that in Figure 6a.

Figure 8c and Figure 8d illustrate the impact of $h/R$ and $k_r$ on the occupied angle $\theta_T$ of the tensile region (as shown in Figure 3a) in the ring system with the nonreciprocal foundation. As indicated in Figure 7b, a lower $k_{rC}/k_{rT}$ produces significant radial deformation in the tensile region on the top to compensate for the reduced resistance in the compressive region on the bottom. In this case, loading is carried by the top region of the elastic foundation, called "top-loading," whose mechanism is widely utilized in pneumatic tires, bicycle wheels, and non-pneumatic wheels [1]. Therefore, the occupied angle of the tensile region, or load-carrying region, in a loaded laminated ring is a valuable indicator for evaluating the "top-loading ratio [36].

A smaller $h/R$ and a greater $k_r (= k_{rT})$ produce a larger $\theta_T$, as shown in Figure 8c and Figure 8d. As shown in Figure 8a and Figure 8b, a larger $\overline{K}_{NF}$ corresponds to sharper deformation at the loading point, resulting in a smaller compressive region ($\theta_C$) and a larger tensile angle ($\theta_T$). Notably, $\overline{K}_{NF} < 1$ produces a minimal sharp local deformation on the loading tip, leading to a larger $\theta_C$ and a smaller $\theta_T$. Based on the analysis presented in Figure 8, we can determine optimal combinations of $h/R$ and $k_r$ to design a laminated ring for a desired load-carrying region while avoiding sharp local deformation.

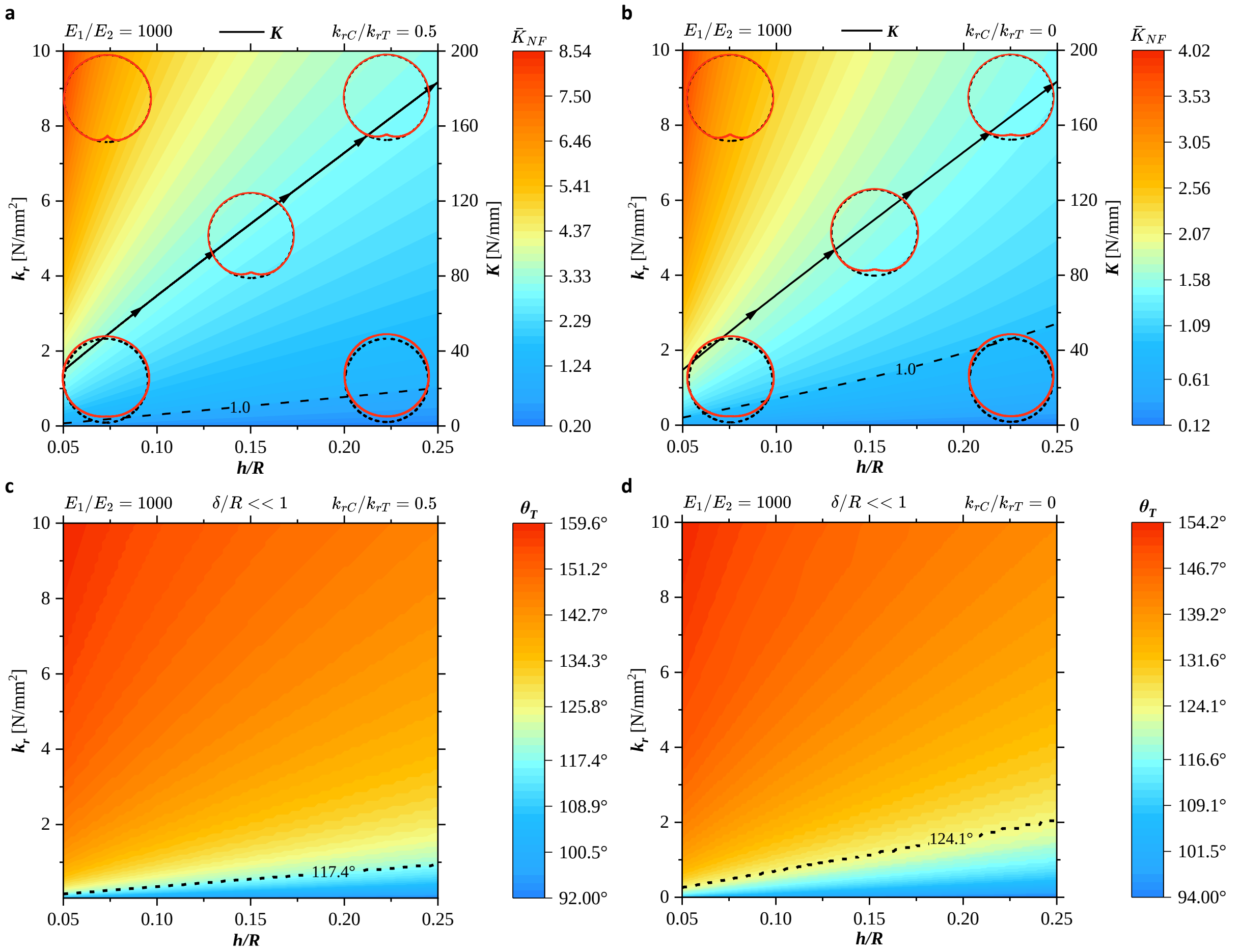

Figure 8. *The normalized stiffness $\bar{K}_{NF}$ shown in (a) and (b), and the occupied tensile angle $\theta_T$shown in (c) and (d) for varying thickness ratio $h/R$ and the nonreciprocal foundation modulus $k_r$ $(= k_{rT})$ where the nonreciprocal ratios are $k_{rC}/k_{rT} = 0.5$ and $k_{rC}/k_{rT} = 0$, respectively.*

### 4.3. Lay-up effect on the laminated ring system

The laminated lay-up can influence the overall ring's deformation mode by interacting with the elastic foundation's stiffness $k_r$. As shown in Figure 9a, a laminated ring can be designed with matrix and fiber-reinforced layers. The matrix's properties are modeled with an isotropic modulus of $E_m = 15$ MPa and Poisson's ratio $\nu_r = 0.48$. The reinforced layers have properties consistent with Section 3.4 and a 0° fiber orientation. We assign geometric parameters such as the overall thickness of the laminated ring $h = 10$ mm, the reinforced layers' thickness at $h_l = 2$ mm, and the mid-layer thickness $h_m$ $(0 \leq h \leq 6\ \mathrm{mm})$. Notably, the exterior matrix's layer $h_e$ is determined by $h_e = (h - 2h_l - h_m)/2$. The foundation modulus $k_r$ varies between 0.01 and 10 N/mm$^2$.

Figure 9b shows the effective stiffness parameters—extension stiffness $(EA)$, bending stiffness $(EI)$, extension-bending coupling stiffness $(EV)$, and shear stiffness $(GA)$—as functions of $h_m$. $EA$ remains constant with changes in $h_m$, primarily influenced by the total cross-sectional area and the material properties of reinforced layers. Notably, $EV$ is a non-zero value due to the curvature of the ring despite the symmetric lay-up. It remains unchanged due to the uniform distribution of reinforcing layers around the neutral axis. In contrast, $EI$ increases with $h_m$ due to the enhanced moment of inertia by the thicker mid-layer positioned farther from the neutral axis, thereby contributing more to the bending resistance. Conversely, $GA$ decreases as $h_m$ increases; a greater mid-layer thickness increases the separation between the reinforced layers, leading to compliance for shear deformation.

Figure 9c and Figure 9d present contour plots of the overall stiffness ($K_{LF}$) of the ring system and $\bar{K}_{LF}$ as functions of $h_m$ and $k_r$. For a soft foundation with a lower $k_r$, $h_m$ does not influence the overall ring's stiffness $K_{LF}$ much, as shown in Figure 9c. As $k_r$ increases, $\bar{K}_{LF}$ rises, enhancing the foundation's resistance to external loadings, which results in greater local deformation of the ring while reducing foundation deformation, as shown in Figure 9d. Consequently, the overall ring system's stiffness $K_{LF}$ becomes more dependent on the ring's stiffness parameters (e.g., $h_m$), as shown in Figure 9c. The black solid line with arrows in Figure 9c indicates optimal combinations for maximum $K_{LF}$. As $k_r$ increases, the optimal $h_m$ value decreases to maximize the $K_{LF}$, as shown inFigure 9c.

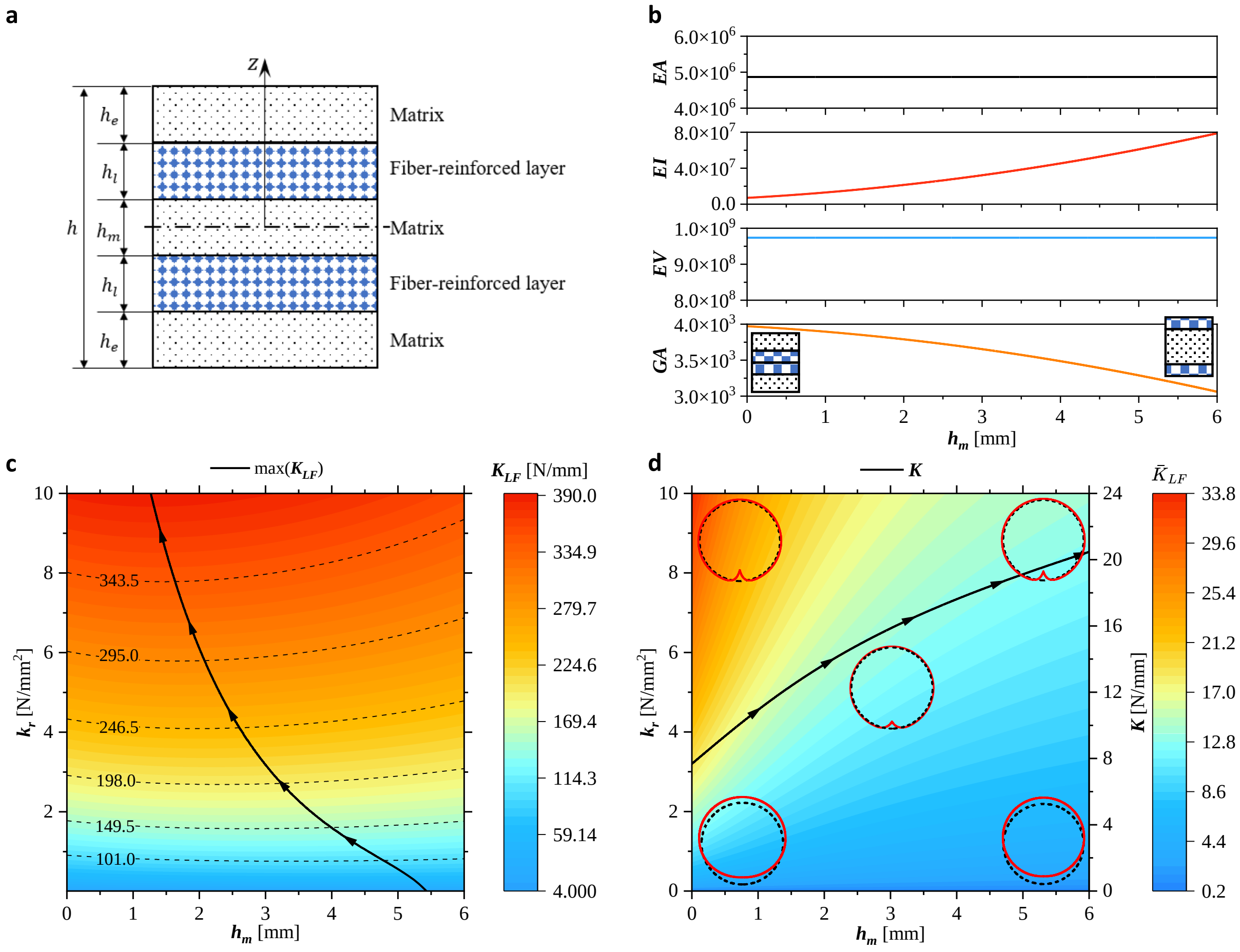

Figure 9. *Lay-up effect on the laminated ring system: (a) Schematic diagram of the cross-section of laminated rings; (b) variations of effective stiffnesses for varying mid-layer thickness* $h_m$*; (c) the ring system's stiffness* $K_{LF}$ *and (d)* $\bar{K}_{LF}$ *and the deformation pattern of laminated rings for varying mid-layer thickness* $h_m$ *and foundation modulus* $k_r$*.*

## 4.4. Response of a laminated ring to nonuniformly distributed loads

Using the scheme outlined in Section 3.2, one can apply a distributed load to a laminated ring on the elastic foundation. For example, we can generate nonuniformly distributed loads using cosine and sine functions within a range of $\theta_F$ in Figure 10a ($-30° \leq \theta_F \leq 30°$) with a maximum magnitude of the load $F = 30$, as shown in Figure 10a and Figure 10b. The laminated ring and elastic foundation parameters are consistent with those in Section 3.4. The deformed configurations under nonuniformly distributed loads clearly show that the maximum radial compressive displacements occur at the location of the load peaks of Cases 1 and 2, as shown in Figure 10b. Unlike a linear elastic foundation, the unilateral nonreciprocal foundation exhibits noticeable radial deformation of the ring system over the loaded regions.

Technically speaking, analytically predicting the deformation of a laminated composite ring on an elastic foundation under nonuniformly distributed loads is a complicated procedure. However, the simplified analytical model for a laminated ring on an elastic foundation, combined with the solution method for a nonreciprocal foundation in Section 3.3, significantly reduces the computational cost, allowing the analysis to be completed within seconds on a modern PC. The MATLAB code of this work is available in Appendix B.

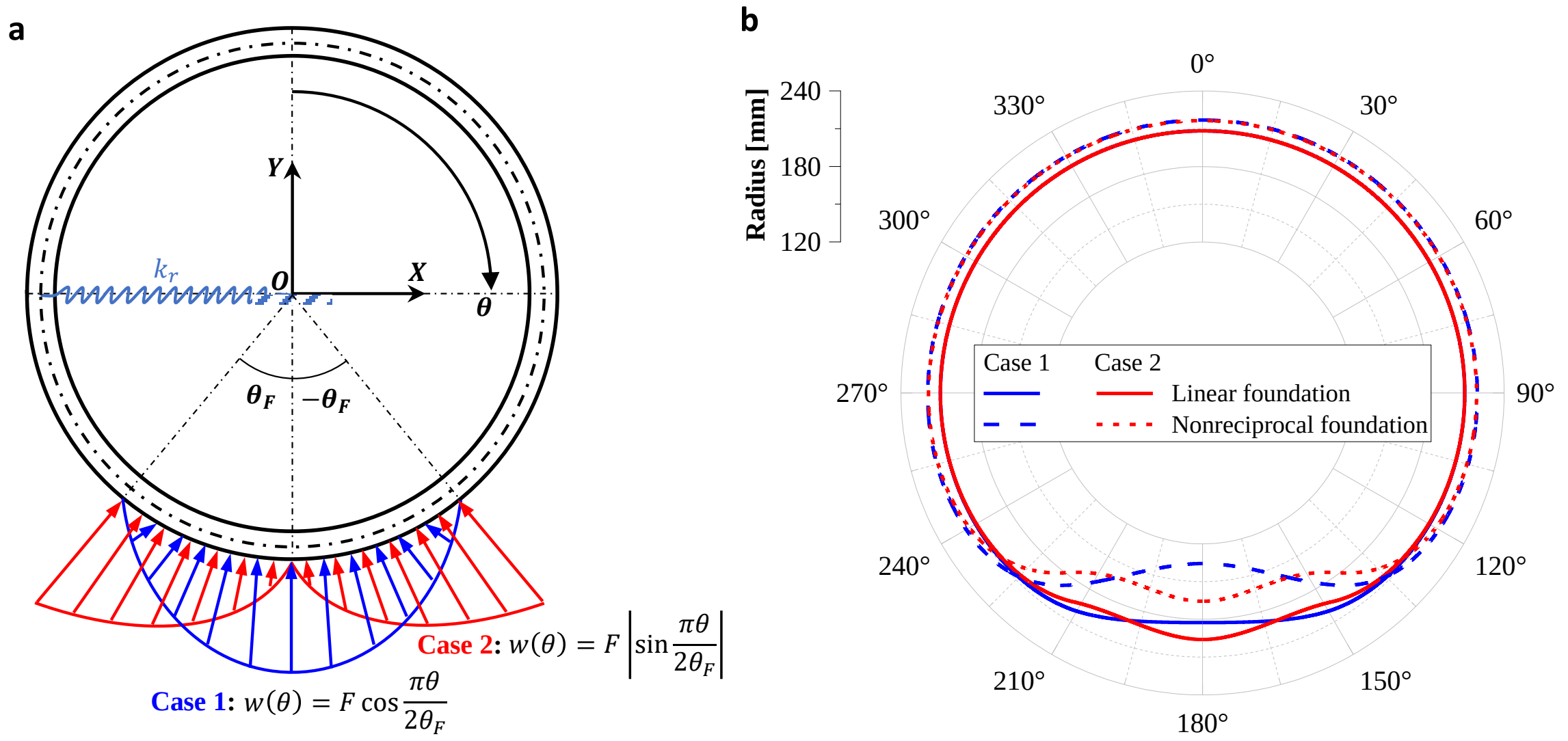

*Figure 10. Responses of the laminated ring subjected to radial distributed loads: (a) loading cases of cosine and sine functions; (b) Deformed patterns of the ring's exterior edge due to the distributed load.*

## 5. Further discussion

The mechanics of laminated rings on elastic foundations under external loads play a critical role in various engineering applications, such as tires, non-pneumatic wheels, bushing bearings, and composite pipes. Despite the need for accurate static modeling to aid in the design of these structures, existing studies predominantly focus on vibration characteristics [27], [28], [29], [30], [43], [44], [45], [46], leaving a gap in the static analysis domain. To bridge this gap, we derive a closed-form solution for a model that accommodates both linear and nonreciprocal foundations under non-axisymmetric concentric and distributed loads. We also quantify the individual contributions of the laminated ring and foundation stiffness to the overall stiffness and deformation of the ring-foundation system, offering valuable design guidelines for their interaction.

Previous works, such as the semi-analytical solution by Gasmi et al. [12] and the closed-form solution by Wang et al. [13], have tackled the problem using Timoshenko beam theory for a homogenized orthotropic ring on an elastic foundation. However, these methods involve complex formulations with numerous parameters. For example, Gasmi et al.'s approach segments the ring into three regions with thirteen interconnected unknowns, while the distributed load due to plate contact complicates calculations of spoke and contact angles, intertwining them with eleven other unknowns and necessitating numerical methods for solving nonlinear equations. Similarly, Wang et al. employed Fourier expansion techniques to decompose radial and tangential displacement solutions into $N$ terms of sine and cosine functions, with each requiring corresponding coefficients. Although this method avoids the direct numerical solutions of the governing equations, achieving high precision requires many terms ($N$), resulting in a proliferation of coefficients.

Our study, in contrast, offers a closed-form solution with only three decoupled unknowns, enabling direct determination of ring deformation and overall stiffness. We extend the closed-form solution to account for complex radially distributed loads by transforming the distributed load into a superposition of radial point loads, thereby avoiding the complexities associated with Fourier expansion [13].

Moreover, existing models assume orthotropic and homogeneous ring behavior [12], [13], which does not accurately reflect the macro-mechanical behavior of laminated rings. In practical applications, such as tires, multiple laminas are bonded together and stacked at different fiber orientations to enhance stiffness and load resistance [47]. To address this, we propose an analytical framework combining the first-order laminated plate theory with a reciprocal elastic foundation, incorporating shear deformation and coupled axial-bending effects. This framework efficiently handles the deformation of laminated rings with various stacking sequences under arbitrary radially distributed load.

While our model simplifies the strain-displacement relationship by neglecting higher-order terms, limiting its accuracy for large deformations (e.g., $\delta/R = 0.15$, as shown in Figure 5a), it closely matches stiffness curves up to $\delta/R = 0.10$. Although the model focuses primarily on laminated rings under radially distributed loads, it does not fully address deformation under tangential loads, which are typically neglected in the spoke design of non-pneumatic tires [1], [36]. Nevertheless, it provides valuable guidance for selecting geometric parameters, material properties for laminated rings, and foundation material moduli. Future work will extend this model to address static contact problems involving arbitrary surface profiles for laminated rings on nonreciprocal foundations.

## 6. Conclusion

In this work, we develop an analytical framework for a laminated composite ring on a nonreciprocal elastic foundation subjected to non-axisymmetric loading. The model generates a design map that correlates the foundation's stiffness with the ring's deformation, considering factors such as ring dimensions, laminate lay-up architecture, and lamina anisotropy. The closed-form solution serves as an efficient design tool for analyzing non-axisymmetric and nonuniform loadings with low computational cost. The design map is valuable for investigating the interaction between the nonreciprocal foundation and the laminated composite ring. The proposed analytical framework and design map have broad applications in engineering fields such as tires, non-pneumatic wheels, flexible bearings, pipelines, expandable tubulars in oil wells, and vascular stents interacting with blood vessel linings, especially under non-axisymmetric loading conditions.

**Acknowledgment**

This research is supported by the National Natural Science Foundation of China (Grant no. 12272225), the Ministry of Science and Technology in China (Grant no. SQ2022YFE010363), and the Research Incentive Program of Recruited Non-Chinese Foreign Faculty by Shanghai Jiao Tong University.

## Appendixes

### A. Laminated ring subjected to a pair of opposite point loads

For the problem shown in Figure A.1, assuming zero distributed loading on the ring ($q_r = q_\theta = 0$), we can decouple Eq. (12) in terms of $u_r$ as

$$\begin{aligned}
&\frac{d^5u_r}{d\theta}+2\frac{d^3u_r}{d\theta}+\frac{du_r}{d\theta}=0\\
&\frac{du_{\theta 0}}{d\theta}=\frac{1}{RA_{11}+B_{11}}\left[\frac{1}{P}\frac{d^4u_r}{d\theta^4}+\left(B_{11}+\frac{1}{P}\right)\frac{d^2u_r}{d\theta^2}\right]-\frac{RA_{11}}{RA_{11}+B_{11}}u_r\\
&\phi=\frac{1}{R}u_{\theta 0}-\frac{1}{R}\left[1+\frac{A_{11}D_{11}-B_{11}^2}{A_{55}(RB_{11}+D_{11})}\right]\frac{du_r}{d\theta}-\frac{A_{11}D_{11}-B_{11}^2}{RA_{55}(RB_{11}+D_{11})}\frac{d^2u_{\theta 0}}{d\theta^2}
\end{aligned}\qquad (A.1)$$

Solving Eq. (A.1) gives the general solutions, which are

$$\begin{aligned}
&u_r(\theta)=C_1\sin\theta+C_2(\cos\theta+\theta\sin\theta)-C_3\cos\theta+C_4(\sin\theta-\theta\cos\theta)+C_5\\
&u_{\theta 0}(\theta)=C_1\cos\theta+C_2\left[\theta\cos\theta-\frac{2(PRA_{11}+1)}{P(RA_{11}+B_{11})}\sin\theta\right]+C_3\sin\theta+\\
&C_4\left[\frac{2(PRA_{11}+1)}{P(RA_{11}+B_{11})}\cos\theta+\theta\sin\theta\right]-C_5\frac{RA_{11}}{RA_{11}+B_{11}}\theta+C_6\\
&\phi(\theta)=-\left[\frac{2A_{55}(RA_{11}+B_{11})}{(A_{11}D_{11}-B_{11}^2)+A_{55}(RB_{11}+D_{11})+RA_{55}(RA_{11}+B_{11})}\right](C_2\sin\theta-C_4\cos\theta)-\\
&C_5\frac{A_{11}\theta}{RA_{11}+B_{11}}+\frac{C_6}{R}
\end{aligned}\qquad (A.2)$$

where $C_i$ $(i = 1, 2, 3, 4, 5, 6)$ are unknown constants to be solved with proper boundary conditions. Considering the symmetry at $\theta = \pi/2$, the boundary conditions are given as follows,

$$u_{\theta 0}(\pi/2)=0,\phi(\pi/2)=0,V(\pi/2)=0 \qquad (A.3)$$

Based on Eq. (A.3), we can rewrite the solutions in Eq. (A.2) as

$$\begin{aligned}
&u_r(\theta)=C_1\sin\theta+C_4\left[\sin\theta-\left(\theta-\frac{\pi}{2}\right)\cos\theta\right]+C_5\\
&u_{\theta 0}(\theta)=C_1\cos\theta+C_4\left[\frac{2(PRA_{11}+1)}{P(RA_{11}+B_{11})}\cos\theta+\left(\theta-\frac{\pi}{2}\right)\sin\theta\right]-C_5\frac{RA_{11}}{RA_{11}+B_{11}}\left(\theta-\frac{\pi}{2}\right)\\
&\phi(\theta)=\frac{2C_4A_{55}(RA_{11}+B_{11})}{(A_{11}D_{11}-B_{11}^2)+A_{55}(RB_{11}+D_{11})+A_{55}(R^2A_{11}+RB_{11})}\cos\theta-\frac{C_5A_{11}}{RA_{11}+B_{11}}\left(\theta-\frac{\pi}{2}\right)
\end{aligned}\qquad (A.4)$$

Besides, due to the symmetry at $\theta = 0$, we can get the following boundary conditions,

$$u_{\theta 0}(0)=0,\phi(0)=0,V(0)=F/2 \qquad (A.5)$$

Solving these boundary conditions provides the three constants adopted in Eq. (A.4) as follows,

$$C_1 = \frac{FR}{2A_{55}}$$
$$C_4 = -\frac{FR[(A_{11}D_{11} - B_{11}^2) + A_{55}(RB_{11} + D_{11}) + RA_{55}(RA_{11} + B_{11})]}{4A_{55}(A_{11}D_{11} - B_{11}^2)} \quad (A.6)$$
$$C_5 = \frac{FR(RA_{11} + B_{11})^2}{\pi A_{11}(A_{11}D_{11} - B_{11}^2)}$$

Therefore, the stiffness $K_{r0}$ of the laminated ring subjected to two opposite point loads can be calculated as

$$K = \frac{F}{|u_r(0)|} = \frac{8\pi}{R\left[\frac{\pi^2}{A_{55}} + \frac{\pi^2(RB_{11} + D_{11})}{A_{11}D_{11} - B_{11}^2} + \frac{R^2A_{11} + RB_{11}}{A_{11}D_{11} - B_{11}^2}\left(\pi^2 - 8 - \frac{8B_{11}}{RA_{11}}\right)\right]} \quad (A.7)$$

Rewriting Eq. (A.7) in terms of the effective stiffnesses in Eq. (40) as

$$K = \frac{8\pi}{\pi^2\frac{R}{GA} + \pi^2\left(\frac{R}{EA} + \frac{R^2}{EV}\right) + \left(\pi^2 - 8 - \frac{8EI}{REV}\right)\left(\frac{R^3}{EI} + \frac{R^2}{EV}\right)} \quad (A.8)$$

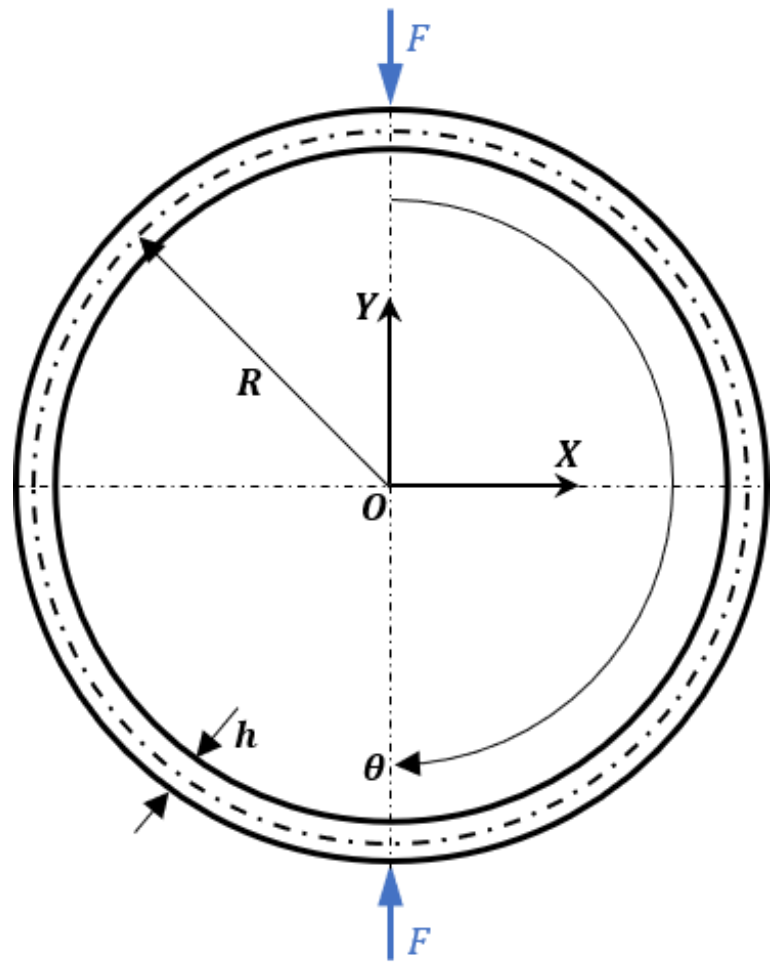

*Figure A.1. A laminated ring subjected to two opposite radial loads, demonstrating the initial conditions and setup for analysis.*

## B. Code availability

MATLAB codes are available on GitHub at https://github.com/Ace628/LREF-model.git